\documentclass[aps,pre,twocolumn,groupedaddress,showpacs,superscriptaddress,amssymb,amsmath,noeprint]{revtex4-2}
\usepackage{graphicx}% Include figure files
\usepackage{dcolumn}% Align table columns on decimal point
\usepackage{bm}% bold math
\usepackage{hyperref}% add hypertext capabilities
\usepackage{cleveref}

\hypersetup{
    colorlinks=true,
    linkcolor=blue,
    urlcolor=blue,
    citecolor=blue
}

\newcommand{\rme}{{\rm e}}
\newcommand{\rmc}{{\rm c}}
\newcommand{\rmd}{{\rm d}}
\newcommand{\rmi}{{\rm i}}

\newcommand{\rmB}{{\rm B}}

\begin{document}
%\newcolumntype{M}[1]{>{\centering\arraybackslash}m{#1}}
%\preprint{APS/123-QED}

\title{Control of spatiotemporal chaos by stochastic resetting}

\author{Camille Aron}
\email{aron@ens.fr} 
\affiliation{Institute of Physics, \'{E}cole Polytechnique F\'{e}d\'{e}rale de Lausanne (EPFL), CH-1015 Lausanne, Switzerland}
\affiliation{Laboratoire de Physique de l’\'{E}cole Normale Sup\'{e}rieure, ENS, Universit\'{e} PSL,
CNRS, Sorbonne Universit\'{e}, Universit\'{e} Paris Cit\'{e}, F-75005 Paris, France}

\author{Manas Kulkarni}
\email{manas.kulkarni@icts.res.in} 
\affiliation{International Centre for Theoretical Sciences, Tata Institute of Fundamental Research,
Bangalore 560089, India}

\date{\today} 

\begin{abstract}
We study how spatiotemporal chaos in dynamical systems can be controlled by stochastically returning them to their initial conditions. Focusing on discrete nonlinear maps, we analyze how key measures of chaos---the Lyapunov exponent and butterfly velocity, which quantify sensitivity to initial perturbations and the ballistic spread of information, respectively---are reduced by stochastic resetting. We identify a critical resetting rate that induces a dynamical phase transition, characterized by the simultaneous vanishing of the Lyapunov exponent and butterfly velocity, effectively arresting the spread of information. These theoretical predictions are validated and illustrated with numerical simulations of the celebrated logistic map and its lattice extension. Beyond discrete maps, our findings are applicable to virtually any chaotic extended classical many-body system.
\end{abstract}

\maketitle

\section{Introduction}\label{sec:introduction}
The spatiotemporal scrambling of information in chaotic many-body systems~\cite{S2014,MSS16,RS16,Bohrdt_2017,SBS16} not only underpins the foundations of statistical physics but also poses substantial practical challenges for the smooth operation of information processing devices as their complexity increases~\cite{CPSB23}. Throughout the history of classical computing, various strategies have been developed to address these challenges and ensure reliable operation within regular dynamical regimes. These include approaches based on feedback and optimal control~\cite{B2003,Zhilin1994,Somdatta1998}, advanced techniques in control theory~\cite{B2000103}, stabilizing periodic orbits via probabilistic control~\cite{ANTONIOU1997,iadecola2022,Pan2024}, and the imposition of dynamical constraints~\cite{2022-deger--lazarides,2022-deger--roy}.
Although this subject has been extensively studied across diverse communities, out-of-time-order correlations (OTOCs) have recently emerged as a relatively transparent probe of spatiotemporal chaos. They bring the well-developed toolbox of correlation functions to the study of dynamical systems. Initially introduced for quantum mechanical systems in the semiclassical regime~\cite{larkin1969,S2014,MSS16,RS16}, OTOCs were subsequently studied in fully quantum systems, such as spin-1/2 chains~\cite{LB2017,NIDS19}, and more recently applied to purely classical systems~\cite{2018-das--bhattacharjee,2021-murugan--ray,2018-bilitweski--moessner,2021-bilitewski--moessner,2020-kumar--dhar,2021-chatterjee--kundu,2020-chatterjee--kulkarni,2018-khemani--nahum,RHK2023,2021-s--kulkarni}.
OTOCs quantify the temporal growth and spatial spread of perturbations between two copies of a system prepared with slightly different initial conditions. 
In classical systems, the exponential sensitivity to initial conditions—a hallmark of chaos—is characterized by Lyapunov exponents, while the ballistic propagation of information is captured by the light-cone structure of OTOCs and quantified by the butterfly velocity.
Alongside ongoing efforts to understand the emergence of these two quantities in diverse classical or quantum setup~\cite{Rozenbaum2017,HeLu2017,Galitski2018,FKPP,rangi2024}, there is a growing interest in devising strategies to control them through externally engineered protocols.

In this work, we explore how stochastic resetting~\cite{EM2011,EM2011a,Evans_2014,KMSS2014,Evans_2020,GJ22} can be harnessed to delay or suppress the onset of spatiotemporal chaos in classical interacting many-body systems.
At its core, stochastic resetting consists in returning the state of a dynamical system to its initial condition at random times.
Variations on this protocol include non-Poissonian resetting times, resetting to a cloud of initial conditions, and more.
Stochastic resetting can also arise naturally in the effective description of deterministic chaotic maps~\cite{Rosa2001,Barkai2009}.
By and large, stochastic resetting has proven to be a versatile nonequilibrium protocol applicable to virtually any dynamical system, with relatively low analytical or computational overhead. It has also been demonstrated experimentally~\cite{Faisant_2021,Roichman2020,Besga2020}.

By introducing a new timescale, stochastic resetting can reshape the system's transient and steady-state dynamics. In the context of random search problems, it can be used to dramatically reduce the mean first-passage time~\cite{Evans_2013,Mercado-Vasquez_2022}.
From a thermodynamic perspective, stochastic resetting has been articulated within the framework of stochastic thermodynamics~\cite{Fuchs_2016,SK23}.
However, much remains to be understood about the impact of stochastic resetting on the dynamics of extended interacting systems~\cite{Nagar_2023}. Studies have begun to explore the effects of resetting on correlations and dynamical phase transitions such as phase synchronization in the Kuramoto model~\cite{bressloff2024kuramoto,SG2022,bressloff2024global}, Kardar-Parisi-Zhang equation~\cite{GMS2014}, exclusion process~\cite{BKP19,Sadekar_2020,Karthika_2020} and the Ising model~\cite{MMS20,Ak20}. Conversely, stochastic resetting can serve as a resource for dynamically generating correlations between non-interacting particles or fields~\cite{BLMS23,BLMS24,KSS2024}.

We question the fate of chaotic properties in interacting many-body dynamics subject to stochastic resetting, within the framework of discrete maps commonly used in the study of dynamical systems~\cite{strogatz2018nonlinear,thompson2002nonlinear,hilborn2000chaos}. After formulating the discrete-time dynamics subject to stochastic resetting~\cite{KMSS2014}, we first focus on Lyapunov growth by analyzing zero-dimensional nonlinear maps subject to resetting. Subsequently, we study the spatiotemporal spread of information in extended systems, particularly the butterfly velocity, using coupled map lattices.

\section{Discrete-time maps under stochastic resetting}\label{sec:discrete_maps}
Let us consider the dynamics generated by a deterministic discrete-time map interspersed with random resetting events. The resulting stochastic map can be formulated as
\begin{align}
\label{eq:gmap}
    x_n \mapsto  x_{n+1} =  \left\{ 
    \begin{array}{ll}
    f(x_n) &\text{with probability } 1-r\\
    x_0 & \text{with probability } r\,,
    \end{array}
    \right.
\end{align}
where $x_n$ denotes the state of the system at time $n \geq 0$.
Here, $0 \leq  r \leq 1$ is the probability, at each time step, to reset the system to its initial condition $x_0$.
Let us first examine how resetting affects the temporal aspect of chaos by considering zero-dimensional maps where $f(x)$ is a non-linear function of the state $x$.
Later, we shall examine the spatial aspects by generalizing to multidimensional-vector states living on a lattice of size $L$, $x_n = \{x_{n,\,i}\}_{i=1 \ldots L}$. The dynamics in Eq.~(\ref{eq:gmap}) are explicitly Markovian but this encompasses non-Markovian evolutions with finite time kernel at the cost of adding auxiliary degrees of freedom.

In the absence of resetting ($r=0$), the deterministic map $f$ is assumed to be ergodic, reaching a unique stationary distribution $p_{\rm st}(x) := \lim\limits_{N\to\infty}\frac{1}{N+1} \sum_{n=0}^N \delta(x-x_n) >0 \ \forall \, x $, independently of the choice of initial condition. 
Additionally, we assume that the deterministic map is chaotic. This manifests as a positive Lyapunov exponent $\lambda$ defined as
$\lambda := \lim\limits_{n\to\infty} \frac{1}{n} \log \big{|} {\delta x_n}/{\delta x_0} \big{|}$, where $\delta x_n$ is the variation of $x_n$ due to an infinitesimal perturbation $\delta x_0$ applied to the initial condition $x_0$. Ergodicity ensures that $\lambda$ is independent of the choice of $x_0$.

In the presence of resetting ($r > 0$), the dynamics in Eq.~(\ref{eq:gmap}) generate stochastic trajectories. The corresponding ensemble average consists in sampling over the many possible realizations of the resetting times.
We show in Appendix~\ref{sec:stat_reset} that the stationary distribution is given by
\begin{align}
\tilde p_{\rm st}(x) = \sum_{n=0}^\infty \tilde g_{n} \, \delta(x-f^n(x_0)) \,,
\end{align}
with $\tilde g_{n} := r(1-r)^n$ and $f^n(x_0)$ stands for the $n$-th iteration of the deterministic map $f(f(\ldots f(x_0)))$.
Although $\tilde p_{\rm st}(x)$ depends on the choice of initial condition $x_0$, none of the results presented below depend on $x_0$ and we do not perform any average over $x_0$.
A rigorous assessment of the ergodicity of the resetting dynamics is challenging and, to the best of our knowledge, has not yet been addressed in the literature~\cite{Aca,Acb,Acc,Acd}. On this specific matter, we shall make informed conjectures rather than definitive conclusions.
To assess its chaotic behavior, \textit{i.e.}, its sensitivity to perturbations of initial conditions, we define the Lyapunov exponent $\tilde \lambda$ as
\begin{align} \label{eq:Lyapunov_def_reset}
      \tilde \lambda := \lim\limits_{n\to\infty}
    \frac{1}{n} \log \, \Big{\langle} \Big{|} \frac{\delta x_n}{\delta x_0} \Big{|} \Big{\rangle}_r \,.
\end{align}
where $\langle \ldots \rangle_r$ denotes the average over resetting times.
Note that, since the dynamics intrinsically comprises both deterministic evolution and stochastic events, the averaging is carried out before taking the logarithm.
Unlike the deterministic case, the independence of  $\tilde \lambda$ with respect to $x_0$ is not trivial and will be clarified below.  
The computation of $\tilde \lambda$ is greatly simplified by expressing $ \tilde d_n  := \langle | {\delta x_n}/{\delta x_0} | \rangle_r $ in Eq.~(\ref{eq:Lyapunov_def_reset}) in terms of its deterministic counterpart: $  d_n  := | {\delta x_n}/{\delta x_0} |$. This is achieved by the renewal formula
\begin{align} \label{eq:renewal}
    \tilde d_n  = \sum_{m = 0}^{n-1} r (1-r)^m d_{m} + (1-r)^n d_n \,,
\end{align}
for $n \geq 1$, together with $ \tilde d_0   =  d_0   = 1 $. Here, the summation on $m$ accounts for averaging over the time elapsed since the last resetting event.

To simplify the derivation of explicit results, we assume that the deterministic dynamics are characterized by a Lyapunov exponent $\lambda$ of order one or larger. Under this assumption, the exponential sensitivity to initial conditions,
 $|\delta x_n / \delta x_0| \propto \exp(\lambda n)$,
develops as early as $n=1$ and we can neglect the sub-exponential contributions. It should be emphasized, however, that the validity of the results is broader, and the restriction to large $\lambda$ will be relaxed later.
With this chaotic ansatz in Eq.~(\ref{eq:renewal}), resetting is found to renormalize the Lyapunov exponent according to 
\begin{align} \label{eq:Lyapunov_resetting}
    \tilde \lambda  = \left\{ 
    \begin{array}{ll}
 \lambda + \log( 1-r) &\text{ if } r < r_\rmc \\
  0 & \text{ if } r \geq r_\rmc \,,
    \end{array}
    \right.
    \end{align}
    where  $r_\rmc := 1- \rme^{-\lambda}$.
This monotonous decay of the Lyapunov with $r$ indicates that increasing the resetting probability systematically reduces the chaoticity of the dynamics.
From Eq.~(\ref{eq:Lyapunov_resetting}), it is now clear that $\tilde \lambda$ is independent of the initial condition $x_0$.
$\tilde \lambda$ vanishes at a critical resetting rate, $r_\rmc\ <1$, signaling a transition from chaotic to non-chaotic dynamics.
For $0< r < r_\rmc$, the dynamics retain exponential sensitivity to initial conditions, and there remains a finite probability for trajectories to explore the entire phase space, preserving ergodicity.
As $r$ approaches $r_\rmc$, the Lyapunov vanishes linearly: $\tilde \lambda  \propto (r_\rmc - r)^\mu$ with the dynamical exponent $\mu = 1$.
At $r = r_\rmc$, a dynamical transition takes place: the exponential sensitivity to initial conditions is lost, and the system shifts to non-chaotic behavior.
For $r \geq r_\rmc$, the dynamics become non-chaotic and we present arguments in Appendix~\ref{sec:ergodicity} indicating a concomitant loss of ergodicity.
In this regime of vanishing Lyapunov, unlike the standard periodic attractor or fixed point pictures, the trajectories are stochastic and remain confined to sequences that revisit the initial state and its early iteration states.
Although generically, a Lyapunov exponent can be negative, Eq.~(\ref{eq:Lyapunov_resetting}) predicts non-negative values. This is because resetting two copies of the map to their respective infinitesimally close initial conditions hinders trajectories from approaching each other exponentially in time.

To numerically test the prediction in Eq.~(\ref{eq:Lyapunov_resetting}), we first reformulate the expression of the Lyapunov exponent in Eq.~(\ref{eq:Lyapunov_def_reset}) using the chain rule $|\delta x_n/\delta x_0| =\prod_{p=0}^{n-1} |f'(x_p)| $ for $n\geq 1$. We obtain an explicit expression that does not rely on any specific assumption regarding $\lambda$,
\begin{align}  \label{eq:cumprod}
      \tilde \lambda = \lim\limits_{n\to\infty}
\frac{1}{n} \log  \Big{[} r & +
r \sum_{m = 1}^{n-1}  (1-r)^{m} \prod_{p=0}^{m-1} |f'(x_p)|  \nonumber \\[-0.5em]
 &+ (1-r)^n \prod_{p=0}^{n-1} |f'(x_p)|  \Big{]}\,.
\end{align}
This approach only requires analyzing a single trajectory of the non-resetting dynamics, thereby avoiding the complex calculations of linear responses to infinitesimal perturbations and the averaging over numerous resetting trajectories.
Equation~(\ref{eq:cumprod}) reveals that $\tilde{\lambda}$ is dominated by those trajectories with a small number of resetting events. This can be seen by inspecting the contributions of the last two terms after substituting into the equation the chaotic ansatz at large $n$, $d_n \propto \exp(\lambda n)$, and using that $d_n$ is at most exponential at early and intermediate times.

\begin{figure}
    \centering
    \includegraphics[width=1\linewidth]{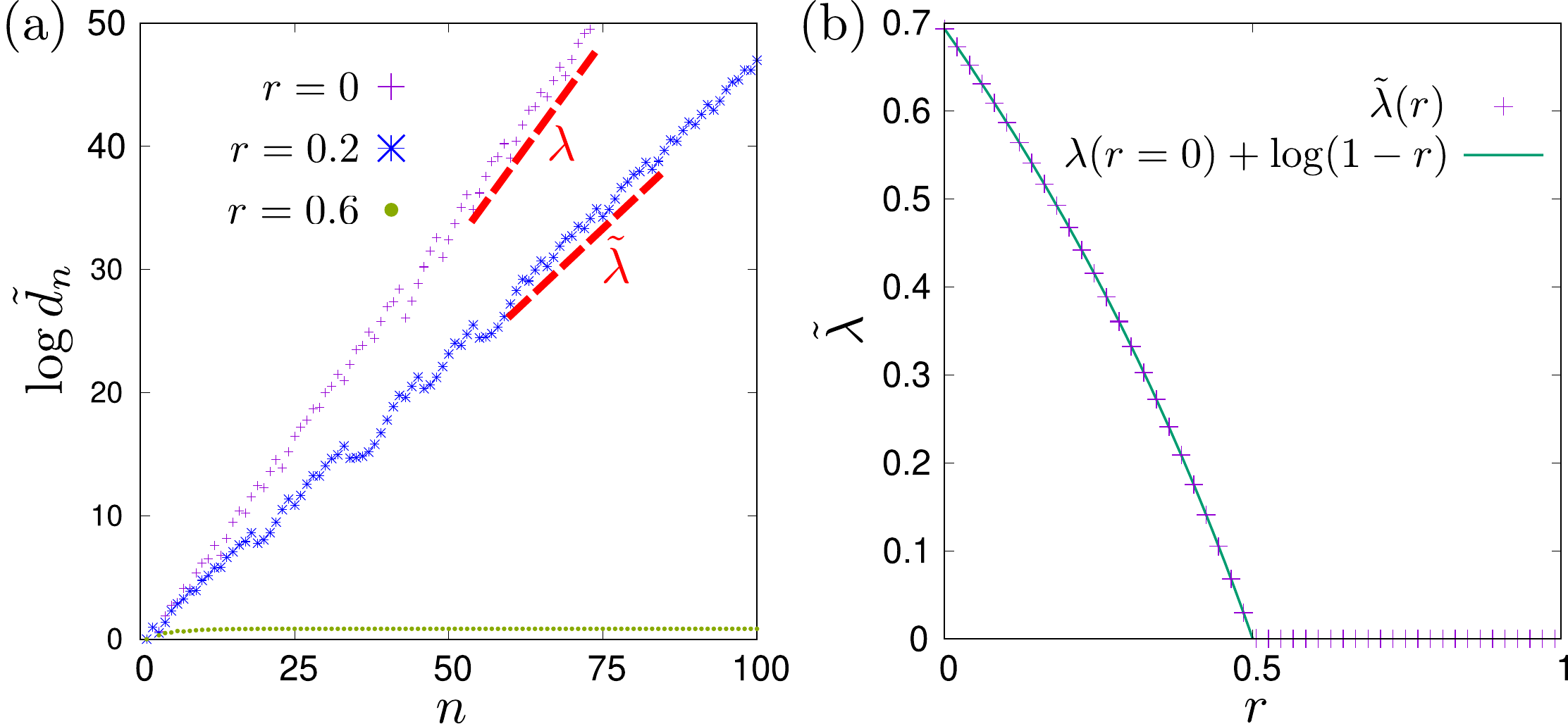}
    \caption{Logistic map [Eq.~(\ref{eq:logistic}), $\alpha=4$] subject to stochastic resetting. 
    (a) Exponential sensitivity to initial perturbations for the deterministic map ($r=0$) and for a resetting rate $r = 0.2 < r_\rmc$.
    The former validates the exponential ansatz $d_ n:=  \log |\delta x_n/\delta x_0| \propto \lambda \,n$, even at early times.
    The Lyapunov exponents $\lambda = \log 2$ and $\tilde \lambda$ can be extracted from the slopes. 
   For $r=0.6 > r_\rmc$, $\tilde d_n$ saturates to a constant.
   (b) The renormalized Lyapunov $\tilde \lambda$ computed with Eq.~(\ref{eq:cumprod}) is compared to the analytical predictions of Eq.~(\ref{eq:Lyapunov_resetting}) as a function of the resetting rate $r$.
   The critical resetting is located at $r_\rmc = 1/2$.
   }
    \label{fig:OTOC_t}
\end{figure}

In practice, we work out the example of the celebrated logistic map which is an archetypal illustration of the onset of chaos~\cite{lorenz1964problem,1976Nature_Robert,mccauley1993chaos, ausloos2006}. 
It is defined as
\begin{align} \label{eq:logistic}
    f(x) = \alpha \, x (1-x)\,,
\end{align}
with the initial state $x_0 \in [0,1]$ and  $\alpha \in [0,4]$ to ensure the stability of the interval $x \in [0,1]$ under the map. 
Let us note that the bifurcation diagram of the logistic map when varying $\alpha \in [0,4]$ remains invariant under stochastic resetting, provided an additional average is performed over initial conditions drawn from the stationary distribution of the deterministic map, see Appendix~\ref{sec:stat_reset}.  
In practice, we work at $\alpha =4$ where the map exhibits fully developed chaos with a Lyapunov $\lambda  = \log 2$ and an ergodic stationary distribution given by $p_{\rm st}(x) = {1}/{(\pi \sqrt{x(1-x)}) }$~\cite{ulam1947mon}.
We emphasize that the analytics leading to Eq.~(\ref{eq:Lyapunov_resetting}) holds when the bare Lyapunov exponent (\textit{i.e.} in the absence of resetting) is positive, and any value of $\alpha$ satisfying this condition will guarantee similar results.
The results are shown in Fig.~\ref{fig:OTOC_t}. Panel~(a) validates the ansatz $d_n \propto \exp(\lambda\, n)$ that was used to derive the renormalized Lyapunov $\tilde \lambda$ in Eq.~(\ref{eq:Lyapunov_resetting}). In Panel~(b), we directly compare these analytical predictions to the numerical results, where Lyapunov exponents are computed using Eq.~(\ref{eq:cumprod}). The agreement is excellent.
When $r > r_\rmc$, the loss of ergodicity is signaled by $\tilde d_n$ plateauing to a finite value which can be interpreted as a quantifier of the volume fraction of the effectively accessible phase space.

\section{Coupled map lattices}\label{sec:coupled_map}
We now examine the spatial aspect of chaos in the presence of stochastic resetting by considering discrete maps on lattices, commonly referred to as coupled map lattices \cite{Kaneko92,KANEKO198960,chazottes2005dynamics}.  
For simplicity, we consider one-dimensional lattices of $L$ sites with periodic boundary conditions.
The state of the system at time $n$ is now represented by a vector $x_n$ with components $x_{n,\,i}$, where $i = 1,,\, \ldots,\, L$.
We follow  both the temporal growth and the spatial spread of a perturbation originating from site $j$ by means of the classical OTOCs defined as
\begin{align} \label{eq:OTOC_lattice}
      D_{n,\,ij} := \Big{|} \frac{\delta x_{n,\,i}}{\delta x_{0,\,j}} \Big{|} \text{ and }  \tilde D_{n,\,ij} := \Big\langle  \Big{|} \frac{\delta x_{n,\,i}}{\delta x_{0,\,j}} \Big{|} \Big\rangle_r\,,
\end{align}
in the deterministic case ($r=0$) and in the presence of resetting ($r>0$), respectively.
To make analytical progress, we assume that the OTOC of the deterministic dynamics takes the form
\begin{align} \label{eq:ansatz_lattice}
    D_{n,\,ij} \propto \exp\left[\lambda \left(\frac{|i-j|}{n} \right)n\right]
\end{align}
for $n \geq |i-j|$, and  $D_{n,\,ij} = 0$ otherwise. Here, the function $\lambda(v)$ is commonly referred to as a velocity-dependent Lyapunov exponent.
This ansatz is known to capture the phenomenology in a wide class of models~\cite{KANEKO1986436,DEISSLER1987397,2018-khemani--nahum,2020-chatterjee--kulkarni}.
We assume that $\lambda(v)$ is a continuous monotonically decreasing function with negative curvature, $\lambda(0) > 0$, and $\lambda(v)$ crosses zero at a finite velocity $v_\rmB$.
At large times, this ansatz describes a traveling wavefront that propagates ballistically at the so-called butterfly velocity $v_\rmB = \lambda^{-1}(0)$, where $\lambda^{-1}$ denotes the inverse function of $\lambda$. At this front, the exponential growth of the OTOC is characterized by the Lyapunov rate $\lambda_\rmB = -v_\rmB \lambda'(v_\rmB) > 0$, see the details in Appendix~\ref{sec:spatial}.
Similarly to Eq.~(\ref{eq:renewal}), the OTOC satisfies the following renewal equation for $n\geq 1$
\begin{align} \label{eq:renewal_lattice}
    \tilde D_{n,\,ij}  = \sum_{m = 0}^{n-1} r (1-r)^m D_{m,\,ij}+ (1-r)^n D_{n,\,ij} \,.
\end{align}
This allows the computation of OTOCs in the presence of stochastic resetting using exclusively data from the deterministic map, thereby avoiding the need for intricate averaging over resetting histories.

In the thermodynamics limit $L\to \infty$ and at large times, we identify a traveling wavefront propagating at the renormalized butterfly velocity  
\begin{align}
    \label{eq:butterfly}
    \tilde v_\rmB =
    \left\{
    \begin{array}{ll}
           \lambda^{-1}\left(-\log(1-r)\right) & \text{ if } r < r_\rmc
     \\
    0 & \text{ if } r \geq r_\rmc\,,
    \end{array}
\right.
\end{align}
where the critical resetting rate is given by $r_\rmc = 1 - \rme^{-\lambda(0)}$.
Without loss of generality, we momentarily simplify the notation by taking the initial perturbation at the site $j=1$ and dropping the $j$ index.
The geometry of the wavefront can be computed by re-expressing the renewal equation in the frame moving at velocity $\tilde v_\rmB$, \textit{i.e.} setting $i = \tilde v_\rmB n + 1+ k$ where $k$ is spatial coordinate in the moving frame. 
When  $\tilde v_\rmB n \gg |k|$, we obtain the front shape
\begin{align}
\lim_{n\to\infty} \tilde D_{n,  \tilde v_{\rm B} n + k
} =  \exp (- k \, \tilde\lambda_{\rmB} / \tilde v_{\rm B})\,.
\end{align}
where the growth rate of the OTOC at the front is characterized by the renormalized Lyapunov exponent
\begin{align} \label{eq:Lyapunov_lattice}
    \tilde \lambda_\rmB = - \tilde v_\rmB  \, \lambda'(\tilde v_\rmB)\,. 
\end{align}
In the phase $r < r_\rmc$, both the butterfly velocity and the Lyapunov are reduced by stochastic resetting.
The near-critical behavior is governed by the behavior of $\lambda(v)$ in the vicinity of $v = 0$. Assuming it is described by $\lambda(v) - \lambda(0) \sim  v^\alpha$ where $\alpha \geq 1$, the renormalized butterfly velocity and Lyapunov exponent scale as $\tilde v_\rmB \sim (r_\rmc -r )^{1/\alpha}$ and $\tilde \lambda_\rmB \sim (r_\rmc -r )^{\mu}$, with $\mu = 1$.
In the phase $r>r_\rmc$, the resetting rate becomes sufficiently strong to drive a dynamical phase transition where the dynamics become non-chaotic, non-ergodic and the spread of information is brought to a complete arrest.
Explicit computations and additional details are provided in the Appendix~\ref{sec:spatial}.

\begin{figure}
\centering
    \includegraphics[width=1\linewidth]{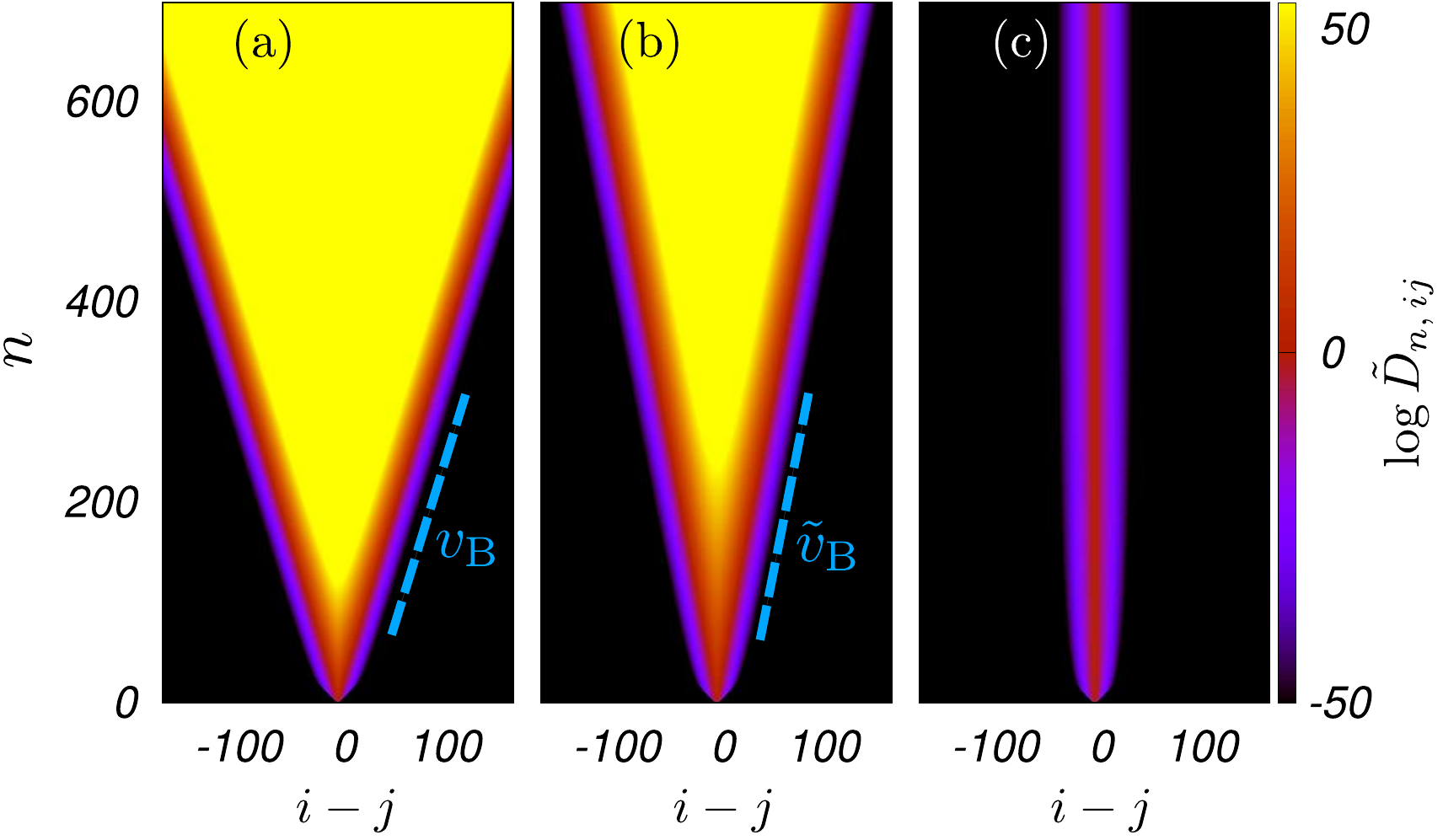}
    \caption{Coupled logistic map [Eq.~(\ref{eq:logistic_lattice}] subject to stochastic resetting: the OTOC $\tilde D_{n,\,ij}$ as a function of space and time reveals the ballistic spreading of perturbations for the
    (a) Deterministic case at $r=0$, and (b) Finite resetting rate $r = 0.2 < r_\rmc \approx 0.38$. (c) Dynamical arrest at $r = 0.4 > r_\rmc$. 
    The butterfly velocities can be extracted from the slopes of the light cones.
    Lattice size $L=701$, $\alpha=4$, and $c=0.1$. The data are spatially averaged over the initial perturbation site $j$, ranging from 1 to $L$.
    \label{fig:lightcone}}
\end{figure}

We illustrate our findings using the coupled logistic map, a one-dimensional lattice of logistic maps interacting with their nearest neighbors. In the absence of resetting, the deterministic dynamics are generated by
\begin{align} \label{eq:logistic_lattice}
x_{n+1,\, i}   &= f(x_{n,\,i}) + \frac{c}{2} \left[ f(x_{n,\,{i-1}}) \!-\!2 f(x_{n,\,i}) \!+\! f(x_{n,\,{i+1}}) \right]\,,
\end{align}
with periodic boundary conditions. The first onsite term is precisely the nonlinearity in Eq.~(\ref{eq:logistic}).
The second term can be seen as a discrete Laplace operator with the diffusive coupling $c \in [0,1]$.
We use a single initial state with random $x_{0,\,i} \in [0,1]$ for $i=1,\ldots,L$. The map preserves $x_{n,\,i} \in [0,1]$ at all times. 
The coupled logistic map exhibits a rich phase diagram and has been widely studied in the context of spatiotemporal chaos, addressing phenomena such as spatial bifurcation, pattern selection, spatiotemporal intermittency, and soliton turbulence~\cite{Kaneko92,KANEKO198960,chazottes2005dynamics,hagerstrom2012experimental,losson1996,MS2022,Notenson2023}. Here, we fix $\alpha = 4$ and $c = 0.1$, corresponding to a regime of fully developed turbulence. Additional details are provided in the Appendix~\ref{sec:lat}.
In the presence of resetting, the OTOCs are numerically obtained with the renewal formula in Eq.~(\ref{eq:renewal_lattice}), making use of the OTOCs of the deterministic map. The latter are computed using an algorithm discussed in the Appendix~\ref{sec:lat}. We restrict times to $2 \tilde v_{\rmB} n < L$ to exclude boundary effects.

\begin{figure}
    \centering
    \includegraphics[width=1\linewidth]{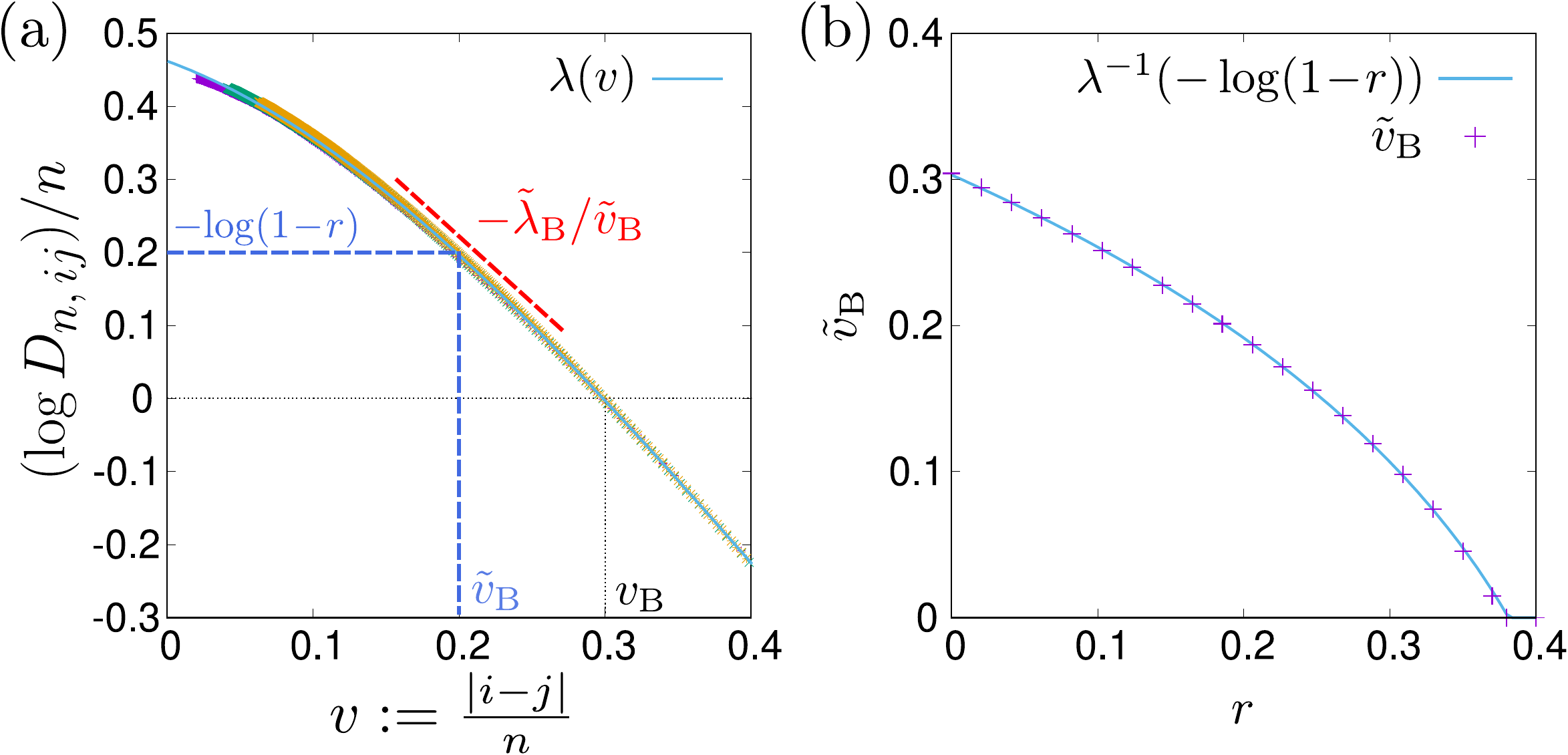}
    \caption{Spatiotemporal chaos in the coupled logistic map subject to stochastic resetting.
 (a) The velocity-dependent Lyapunov $\lambda(v)$ [Eq.~(\ref{eq:ansatz_lattice})] is numerically extracted from the collapse of $(\log D_{n,\, ij}) / n $ versus $v:=|i-j|/n$ for various values of $i-j$ in the deterministic case ($r=0$).
 The butterfly velocity $\tilde v_\rmB$ at $r>0$ can be determined by solving $\lambda(\tilde v_\rmB) = -\log(1-r)$. The slope of $\lambda(v)$ at this solution determines the Lyapunov rate $\tilde \lambda_\rmB$. 
 (b) $\tilde v_\rmB$ is extracted from the slope of the light cones (see Fig.~\ref{fig:lightcone}) and is compared to the prediction of Eq.~(\ref{eq:butterfly}) as a function of the resetting rate $r$. Lattice size $L=701$, $\alpha = 4$, and $c = 0.1$.}
    \label{fig:collapse}
\end{figure}

Figure~\ref{fig:lightcone} illustrates the spatiotemporal behavior of the OTOC $\tilde D_{n,\,ij}$ under different resetting conditions.
Panel (a) shows the light-cone structure of the OTOC in the absence of resetting ($r=0$).
The local perturbation of the initial condition evolves into a traveling wavefront that propagates ballistically at the butterfly velocity $v_\rmB$ visible as the slope of the light cone at late times. 
Note that the causal light cone at the maximal Lieb-Robinson velocity $v_{\rm LR} = 1$ can be seen at early times, and it is overshadowed by the prominence of the butterfly front at later times.
In Panel (b),  the butterfly velocity is visibly reduced by a finite resetting rate ($r = 0.2$).
In Panel (c), for $r > r_\rmc$, 
the OTOC quickly saturates to a function $\tilde D_{ \infty,\,ij} := \lim_{n\to\infty} \tilde D_{n,\,ij} $ that is exponentially localized around the perturbation site $j$.
The light cone collapses as the dynamics freeze, signaling the arrest of spatiotemporal chaos.
Noteworthy, a similar dynamical phase transition was reported in deterministic classical many-body systems with kinetic constraints~\cite{2022-deger--lazarides,2022-deger--roy}.

In Fig.~\ref{fig:collapse}, we test the theoretical predictions of Eq.~(\ref{eq:butterfly}) for the reduction of the butterfly velocity by stochastic resetting.  
In Panel (a), the velocity-dependent Lyapunov exponent $\lambda(v)$ is extracted by collapsing the OTOC data at $r=0$ onto a single master curve, validating the ansatz in Eq.~(\ref{eq:ansatz_lattice}).  
In Panel (b), we extract the butterfly velocities in the presence of resetting from the analysis of the light cones and compare them with the theoretical predictions of Eq.~(\ref{eq:butterfly}). The agreement between the numerical results and the analytical predictions is excellent.

\section{Discussions and outlook} \label{sec:discussion}
We have shown that stochastic resetting provides a powerful mechanism for mitigating spatiotemporal chaos and can even induce a dynamical phase transition to a non-chaotic regime, where the spreading of information is brought to a complete halt.
This can be put in perspective with the quantum many-body localized regimes where, although slower than ballistic, the spreading of information is thought to survive in the absence of transport.
More generally, the insights gained from these classical results will pave the path to understanding ergodicity and chaos in many-body quantum systems subject to resetting \cite{MSM18,RT18,MC22,KM23}.
While we focused on discrete nonlinear maps, our results readily adapt to continuous-time dynamics. Using the dictionary $\lambda \to \lambda \, \rmd t$ and $r \to r \,\rmd t$, the renormalized Lyapunov in Eq.~(\ref{eq:Lyapunov_resetting}) becomes $\tilde \lambda = \lambda - r$, the butterfly velocity in Eq.~(\ref{eq:butterfly}) becomes $\tilde v_\rmB = \lambda^{-1}(r)$, and the expression of $\tilde \lambda_\rmB$  in (\ref{eq:Lyapunov_lattice}) remains valid, see Appendix~\ref{sec:ctd} for more details.
The versatility of these results suggests the possibility of developing a geometrical optics theory to investigate phenomena such as the refraction of ``information light rays'' on spatial domains subject to resetting.

\section*{Acknowledgements} M.K. acknowledges support from the Department of Atomic Energy, Government of India, under project No. RTI4001. M.K. also thanks the VAJRA faculty scheme (No. VJR/2019/000079) from the Science and Engineering Research Board (SERB), Department of Science and Technology, Government of India. We gratefully acknowledge support from the International Research Project (IRP) titled `Classical and Quantum Dynamics in Out of Equilibrium Systems' (DynoutSys), funded by CNRS, France. M.K. thanks the Institute of Physics at EPFL in Lausanne for their hospitality. C.A. thanks the International Centre for Theoretical Sciences (ICTS) in Bangalore for their hospitality.
This research was supported in part by the ICTS and CEFIPRA for participating in the ``Indo-French Workshop on Classical and quantum dynamics in out of equilibrium systems'' (ICTS/IFWCQM2024/12).

%\onecolumngrid

%\setcounter{equation}{0}
%\setcounter{figure}{0}
%\renewcommand{\theequation}{S\arabic{equation}}
%\renewcommand{\thefigure}{S\arabic{figure}}

%\newpage
\appendix

\section{Discrete-time maps subject to stochastic resetting}
\label{sec:dtm}
\subsection{Stationary distribution of the deterministic map} \label{sec:stat}
Let us first briefly discuss the deterministic case, in the absence of resetting ($r=0$). We consider a zero-dimensional map of the form 
\begin{align}
    x_n \mapsto  x_{n+1} = f(x_n) \,,
\end{align}
where $f(x)$ is a nonlinear function of $x$ and $n\geq 0$.
We assume that the deterministic map has a unique stationary distribution $p_{\rm st}(x)$ and that the dynamics are ergodic. The ergodic theorem ensures that 
\begin{align} \label{eq:p_st}
p_{\rm st}(x) = \lim\limits_{N\to\infty}\frac{1}{N+1} \sum_{n=0}^N \delta(x-x_n)
\end{align}
and $p_{\rm st}(x) > 0$ for all $x$, \textit{i.e.}, any region of phase space will be visited with finite frequency. Note that the expression above is ``democratic'' in the sense that it is invariant under any permutation of the $x_n$'s.

\subsection{Stationary distribution with stochastic resetting}\label{sec:stat_reset}
In the presence of a finite resetting probability ($0< r <1$), the dynamics are generated by the following stochastic map
\begin{align}
    x_n \mapsto  x_{n+1} =  \left\{ 
    \begin{array}{ll}
    f(x_n) &\text{with probability } 1-r\\
    x_0 & \text{with probability } r\,.
    \end{array}
    \right.
\end{align}
The corresponding stationary distribution, if it exists, is given by renewal equation 
\begin{align} \label{eq:p_st_resetting}
\tilde p_{\rm st}(x) = \sum_{m=0}^\infty \tilde g_m \, \delta(x-f^m(x_0))\,,
\end{align}
where  $\tilde g_m := r(1-r)^m$ is the probability that a reset occurred exactly 
$m$ time steps ago, and that no reset has occurred in the subsequent 
$m$ steps. One may check the normalization
\begin{align}
    \int \rmd x \, \tilde p_{\rm st}(x) = \sum_{m=0}^\infty  \tilde g_m = 1\,.
\end{align}
Resetting is naturally expected to enhance the overall time spent by the system in the initial state, $x_0$, and its first subsequent images under the map $f$. This is confirmed by examining Eq.~(\ref{eq:p_st_resetting}), which, unlike the deterministic case in Eq.~(\ref{eq:p_st}), lacks invariance under arbitrary permutations of the $x_n$'s.
Notably, the initial condition exhibits the highest weight to the stationary distribution $\tilde p_{\rm st}(x)$, given by $\tilde g_0 = r$. This demonstrates the dependence of the stationary distribution $\tilde p_{\rm st}(x)$ on the choice of the initial condition.

If, additionally, one averages the resetting dynamics over the initial conditions by sampling $x_0$ from the stationary distribution of the deterministic map, \textit{i.e.}, with probability $p_{\rm st}(x_0)$, one can show by induction that $p_{\rm st}(x)$ remains an invariant distribution of the resetting dynamics.
Assume that $x_n$ is sampled from $p_{\rm st}$. 
\begin{itemize}
    \item If there is no resetting at time $n+1$, then $x_{n+1} = f(x_n)$ is also sampled from $p_{\rm st}$ since $p_{\rm st}$ is precisely the invariant distribution of the deterministic map.
    \item If there is a resetting event at time $n+1$, then $x_{n+1} = x_0$, which is also sampled from $p_{\rm st}$ by construction.
\end{itemize}
In both cases, $x_{n+1}$ is sampled from $p_{\rm st}$. 
Since this holds true for $n=0$ (by the initial sampling of $x_0$), it holds true for all times $n \geq 0$ by induction.

This result states that, under such an averaging procedure of the initial conditions, the statics of the map are not altered by stochastic resetting. For instance, this implies that the celebrated bifurcation diagram of the logistic map is not altered by stochastic resetting.
However, in this work, we do not perform this additional averaging over the initial conditions. Instead, we focus on the case of a single, arbitrary, initial condition $x_0$. We have carefully checked that our results do not depend on the choice of $x_0$.

\subsection{Ergodicity with stochastic resetting} \label{sec:ergodicity}
Here, we provide semi-rigorous arguments to assess the ergodicity of the dynamics subject to stochastic resetting, assuming that the deterministic dynamics are ergodic in a bounded phase space and chaotic with a finite Lyapunov exponent $\lambda > 0$. We caution that the results presented below should only be regarded as indicative until this problem receives a fully rigorous analysis.

By ergodicity, we mean that for almost all initial conditions $x_0$, the subsequent system's trajectory passes arbitrarily close to any point in phase space, ensuring that the system eventually explores all regions of phase space with a frequency proportional to their volume. Note that this definition does not necessarily require independence (of these frequencies) with respect to initial conditions. More precisely, the ergodicity of the deterministic map, 
means that almost all $x$ in phase space, a given small $\epsilon >0$, and a large enough time $N$, the condition $|x-x_n| < \epsilon$ is met by a finite set $\Lambda_{\epsilon,N}(x)$ of $n$'s along the trajectory before time $N$ is reached. The size of this set is on the order of $\mathcal{O}(\epsilon\times N)$, and ergodicity simply implies the nonvanishing of the stationary distribution, $p_{\rm st}(x) > 0$, for almost all $x$.
In practice, we question whether the stationary distribution in the presence of stochastic resetting is non-vanishing for arbitrary generic values of $x$, \textit{i.e.} $\tilde p_{\rm st}(x) > 0$.
Integrating Eq.~(\ref{eq:p_st_resetting}) on a small interval of length $2 \epsilon$ around $x$, we get
\begin{align}
\int_{x-\epsilon}^{x+\epsilon} \!\!\!\! \rmd y \, \tilde p_{\rm st}(y) = \sum_{n \in \Lambda_{\epsilon}(y)} \tilde g_{n} \,,
\end{align}
where the infinite set that appears in the summation $\Lambda_{\epsilon}(x) := \lim\limits_{N\to\infty} \Lambda_{\epsilon,N}(x)$ collects all the times at which a given trajectory, initialized at $x_0$, comes at a distance $\epsilon$ of the target $x$.
For $\epsilon$ small enough, this yields
\begin{align} \label{eq:sumpo}
\tilde p_{\rm st}(x) = \frac{1}{2\epsilon} \sum_{n \in \Lambda_{\epsilon
}(x)} \tilde g_{n} \,.
\end{align}
All the $g_n$'s being strictly positive, we may therefore find a lower bound to the sum over the infinite set $\Lambda_{\epsilon}(x)$ by keeping a single term only:
\begin{align}
\tilde p_{\rm st}(x) > \frac{1}{2\epsilon} \tilde g_{n_{\rm min}} \,,
\end{align}
where $n_{\rm min}$ is chosen to be the smallest element of $\Lambda_{\epsilon}(x)$, corresponding to the largest weight $\tilde g_{n}$ of the sum in Eq.~(\ref{eq:sumpo}). We now use the chaotic character of the deterministic dynamics to get an estimate for $n_{\rm min}$: the typical distance between two trajectories originating from neighboring initial conditions at a small enough distance $\epsilon$ grows exponentially
\begin{align} 
\frac{|\delta x_n|}{\epsilon} \sim \rme^{\lambda n} \text{ at large enough $n$,}
\end{align}
with the Lyapunov exponent $\lambda > 0$.
Identifying $|\delta x_n|$ with the typical size of phase space (assumed to be bounded), the waiting time of a time-reversed trajectory to come at a close distance $\epsilon$ of the target $x$ can be estimated as
\begin{align}
n_{\rm min} \sim \frac{1}{\lambda} \log \frac{1}{\epsilon}\,.
\end{align}
Using the expression of $\tilde g_n$, this yields the bound
\begin{align}
\tilde p_{\rm st}(x) \gtrsim \frac{r}{2}  \, \left(\frac{1}{\epsilon}\right)^{\frac{\log (1-r)}{\lambda}+1}\,.
\end{align}
Inspection of the right-hand side of the above inequality indicates a drastic change in the ergodic nature when the exponent $\frac{\log (1-r)}{\lambda} + 1$ vanishes. This occurs precisely at the critical resetting rate $r_\mathrm{c} = \mathrm{e}^{-\lambda} - 1$, where the renormalized Lyapunov exponent $\tilde \lambda$ vanishes.
The remarkable concurrence of these two transitions leads us to propose the following conjecture:
For $r < r_\mathrm{c}$, the dynamics are ergodic, whereas ergodicity is lost for $r > r_\mathrm{c}$.

\section{Deterministic coupled logistic map}
\label{sec:lat}
In this Section, we briefly present the coupled logistic map in the absence of stochastic resetting ($r=0$) and explain how to efficiently compute the corresponding OTOC $D_{n,\,ij}$ introduced in Eq.~(\ref{eq:OTOC_lattice}) of the main manuscript, circumventing the delicate numerical computation of linear responses to infinitesimal perturbations. The OTOC in the presence of resetting, $\tilde D_{n,\,ij}$, is obtained by means of the renewal formula in Eq.~(\ref{eq:renewal_lattice}) which circumvents the delicate average over resetting trajectories.

\begin{figure}
    \centering
    \includegraphics[width=0.6\linewidth]{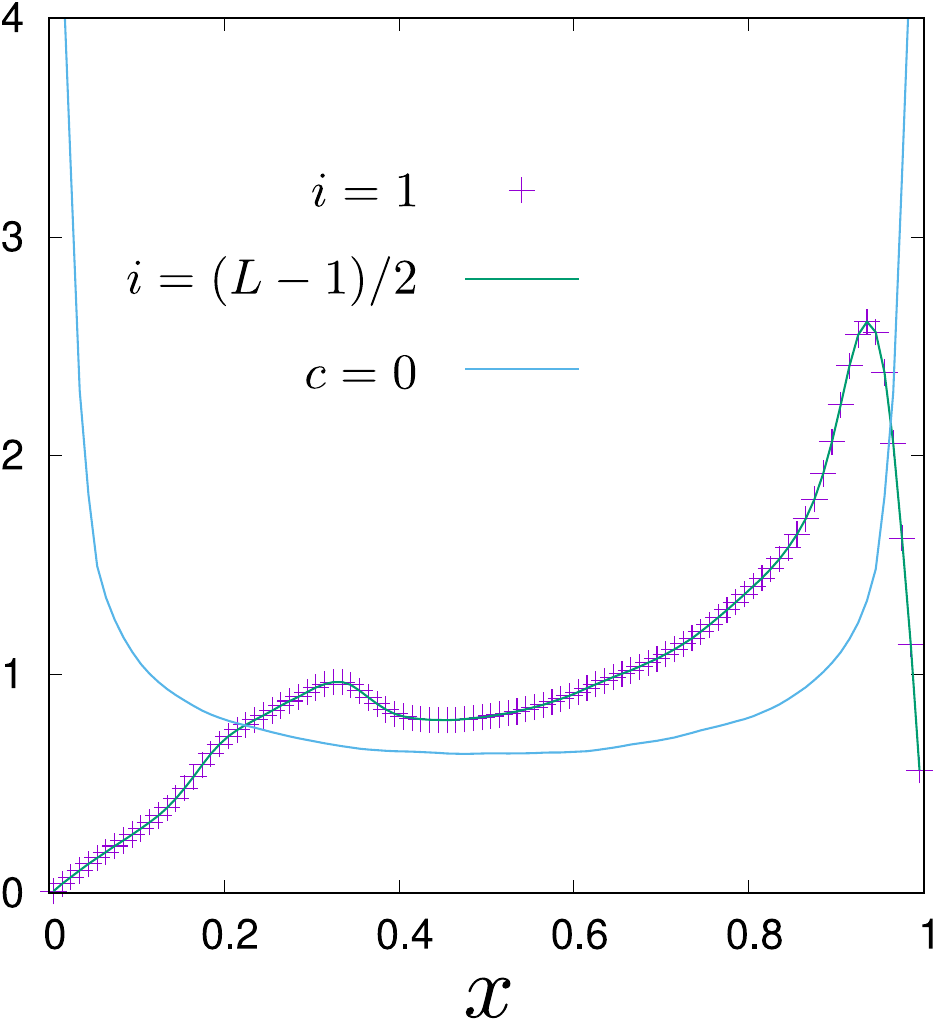}
    \caption{
    Coupled logistic map with $\alpha=4$ and $c=0.1$ (no resetting, $r=0$): the local marginal of the stationary distribution is identical at all sites, here $i=1$ and $i=(L-1)/2$. The stationary distribution of the corresponding logistic map ($c=0$),  $p_{\rm st}(x)
= 1/(\pi \sqrt{x(1-x)})$, is plotted for comparison.
    Lattice size $L=701$.
    }
    \label{fig:distribution}
\end{figure}

The coupled logistic map consists of a one-dimensional chain of $L$ sites where each site hosts a single logistic map that interacts diffusively with its nearest neighbors. At time $n=0$, the state of the system is specified by the vector $x_0 = \{x_{0,\,i} \}_{i=1\ldots L}$ with $|x_{0,\,i}| \leq 1$.
The deterministic dynamics are generated by
\begin{align}
x_{n+1,\, i} = F_i(x_n)\,,
\end{align}
with the nonlinear vector map 
\begin{align}
F_i(x_n) =   f(x_{n,\,i}) + \frac{c}{2} \left[ f(x_{n,\,{i-1}}) -2 f(x_{n,\,i}) + f(x_{n,\,{i+1}}) \right] \,,
\end{align}
together with periodic boundary conditions. Here, $f(x) = \alpha \, x(1-x)$ is the logistic function with the parameter $\alpha \in [0,4]$ and $0< c < 1$ is a diffusion parameter (the term in brackets is a discrete version of the one-dimensional Laplacian). This ensures the multidimensional interval $[0,1]^L$ is stable under the coupled logistic map. 
In practice, we work with $\alpha=4$ and $c=0.1$ for which we have carefully checked that the deterministic dynamics reach a stationary state that is both uniform in space and time.
This is illustrated in Fig.~\ref{fig:distribution} where the local marginal of the stationary distribution is plotted at sites $i=1$ and $i = (L-1)/2$. For comparison, we also plot $p_{\rm st}(x)
= 1/(\pi \sqrt{x(1-x)})$ which is the stationary distribution of the logistic map (at $c=0$).

The OTOC of the deterministic dynamics is defined as
\begin{align}
    D_{n,\,ij} = \left| \frac{\delta x_{n,\,i}}{\delta x_{0,\,j}} \right|\,.
\end{align}
At $n = 0$, this is simply $D_{0,\,ij} = \delta_{ij}$. For $n \geq 1$, the chain rule offers a convenient representation as a matrix product
\begin{align} \label{eq:cumprod_lattice}
    D_{n,\,ij} =  | \left\{F'(x_0) 
    \times F'(x_1) \times \ldots \times F'(x_{n-1}) \right\}_{ij} | \,,
\end{align}
where we introduced the Jacobian matrix 
\begin{align}
    F_{ij}'(x_n) := \frac{\partial F_i(x_n) }{\partial x_{n,\, j} }\,,
\end{align}
and $\times$ denotes matrix multiplication in the spatial indices. In the particular case of the logistic lattice, we have
\begin{align}
    F_{ij}'(x) =&  (1-c) \frac{\partial f(x_i)}{\partial x_j} \delta_{ij} \nonumber \\
    &+\frac{c}{2} \! \left[\frac{\partial f(x_i)}{\partial x_{j+1}} \delta_{i,j+1} + \frac{\partial f(x_i)}{\partial x_{j-1}} \delta_{i,j-1}\right]\,,
\end{align}
together with periodic boundary conditions.

\section{Continuous-time dynamics subject to stochastic resetting} \label{sec:ctd}
In this Section, we formulate the resetting dynamics for dynamical systems with a continuous-time description. Although very similar to the discrete-time case, we find the continuous-time case more amenable for presenting analytical proofs.

\subsection{Zero-dimensional systems (no space)}\label{sec:dico}
To simplify the discussion, we use similar notations and the following dictionary between the discrete-time and the continuous-time formulations:
\begin{align}
    \begin{array}{rl}
         n & \longleftrightarrow t/\rmd t\\
    x_n & \longleftrightarrow x(t) \\
     x_{n+1} - x_n& \longleftrightarrow \partial_t x(t)
         \end{array}
         \qquad
         \begin{array}{rl}
     \lambda & \longleftrightarrow \lambda \, \rmd t \\
    r &\longleftrightarrow r \, \rmd t \\
    (1-r)^n & \longleftrightarrow \rme^{-r t}
    \end{array}
\end{align}
where $\lambda$ and $r$ are now rates instead of dimensionless exponent and probability, respectively. In particular, the resetting rate $r$ now takes values in $[0, \infty)$. 

\subsubsection{Deterministic dynamics \&  Lyapunov rate}
Let us consider deterministic dynamics governed by the following first-order PDE at times $t > 0$
\begin{align}
\partial_t x(t) = f(x(t))\,,
\end{align}
where $f(x)$ is a non-linear function of $x$ which is assumed to generate ergodic and chaotic dynamics, and the initial condition $x(0)$ at $t=0$.

The absolute distance between two trajectories starting from neighboring initial conditions is defined as
\begin{align}\label{eq:def_d}
d(t) := \left| \frac{\delta x(t)}{\delta x(0)} \right|\,.
\end{align}
To make the connection with the study of quantum chaos, the quantity $d(t)$ can be referred to as an ``out-of-time-order correlator'' (OTOC). Indeed, in a semi-classical sense, $ {\delta x(t)}/{\delta x(0)} $ is to be replaced by the commutator $[\hat x(t), \hat p(0)]/\rmi\hbar$ where $\hat p$ is the momentum operator conjugated with the position operator $\hat x$, $[\hat x, \hat p] = \rmi \hbar$. Hence, $d^2(t)$ is to be replaced by the square commutator $\langle [\hat x(t), \hat p(0)]^\dagger [\hat x(t), \hat p(0)] \rangle/(\rmi\hbar)^2$ where quantum mechanical operators explicitly appear out of time order and $\langle \ldots \rangle$ denotes averaging over an appropriately chosen initial density matrix.

The linearization of the deterministic dynamics for two close-by trajectories distant of $\delta x(t)$ yields the first-order PDE
\begin{align}
\partial_t  \delta x(t) = \delta x(t) f'(x(t))\,,
\end{align}
which is solved as
\begin{align} \label{eq:d_solution}
d(t) = \rme^{\int_0^t \rmd \tau |f'(x(\tau))| }\,.
\end{align}
The ergodic theorem notably ensures
\begin{align} \label{eq:ergo}
\lim\limits_{t \to\infty}\frac{1}{t} \int_0^t \rmd \tau \, |f'(x(\tau))| =  \int\rmd x \, p_{\rm st}(x) |f'(x)| =: \lambda > 0\,.
\end{align}
which provides a static statistical expression of the Lyapunov rate. It connects to the traditional dynamical definition of the Lyapunov via the asymptotic expression at large times $t \gg 1/\lambda$
\begin{align}
d(t) \simeq \rme^{\lambda t}\,.
\end{align}

\medskip

\subsubsection{Dynamics subject to stochastic resetting \& Lyapunov rate}
In the presence of stochastic resetting, the dynamics are generated by
\begin{widetext}
\begin{align}
x(t) \mapsto
x(t+\rmd t) = \left\{ 
\begin{array}{ll}
      x(t) + f(x(t)) \, \rmd t  &\text{ with probability $1-r\, \rmd t$} \\
  x(0)  & \text{ with probability $r\, \rmd t$} \,.
\end{array}
\right.
\end{align}
\end{widetext}
We consider the OTOC
\begin{align} \label{eq:def_d_tilde}
\tilde d(t) := \Big{\langle} \left| \frac{\delta x(t)}{\delta x(0)} \right| \Big{\rangle}_r\,,
\end{align}
where $\langle \ldots \rangle_r$ denotes the average with respect to the resetting events.
$\tilde d(t)$ can be simply related to its deterministic counterpart $d(t)$ in Eq.~(\ref{eq:def_d}) via the following renewal formula
\begin{align} \label{eq:renewal_cont}
\tilde d(t)  = r \int_0^t \rmd \tau \, \rme^{-r \tau} d(\tau) + \rme^{-r t} d(t)\,,
\end{align}
which is the continuous-time version of Eq.~(\ref{eq:renewal}) in the main part of the manuscript.
Using the solution in Eq.~(\ref{eq:d_solution}), the renewal formula yields
\begin{align} \label{eq:bnji}
\tilde d(t) = r \int_0^t \rmd \tau \, \rme^{-r\tau + \int_0^\tau \rmd u \, |f'(x(u))|} +  \rme^{-rt + \int_0^t \rmd \tau \, |f'(x(\tau))|   } \,.
\end{align}
Let us now divide the first time integral in Eq.~(\ref{eq:bnji}) as
\begin{align}
\int_0^t \rmd \tau = \int_0^{z(t)} \rmd \tau + \int_{z(t)}^t \rmd \tau\,,
\end{align}
where we introduced the \textit{ad-hoc} timescale $z(t) \gg 1/\lambda$ that is chosen to be monotonously growing slower than any power-law growth, \textit{i.e.} $\lim\limits_{t\to\infty} z(t)/t^\alpha =0 \ \forall \, \alpha >0$. Working at large time $t$, using the ergodic theorem in Eq.~(\ref{eq:ergo}) in the large time window $[z(t),t]$, we obtain the asymptotic expression
\begin{align} \label{eq:d_efg}
\tilde d(t) \simeq & \ r \int_0^{z(t)} \rmd \tau \, \rme^{-r\tau + \int_0^\tau \rmd u \, |f'(x(u))|} \nonumber \\
&+ \frac{r}{\lambda-r}   \left[ \rme^{(\lambda-r) t} - \rme^{( \lambda-r ) z(t)}  \right] + \rme^{(\lambda-r) t} \,.
\end{align}
This asymptotic expression allows us to distinguish two dynamical regimes depending on the relative strength of the resetting rate $r$ with the critical rate $r_\rmc = \lambda$.
If $r < r_\rmc  $, those of the terms above that involve $z(t)$ are subdominant, and we obtain the exponential growth
\begin{align} \label{eq:d_tilde}
\tilde d(t)  \simeq \left( 1 + \frac{r}{\tilde \lambda} \right) \rme^{ \tilde \lambda t}\,,
\end{align}
where the renormalized Lyapunov rate is given by
\begin{align}
\tilde \lambda = \lambda - r > 0 \text{ if } r < r_\rmc\,.
\end{align}
Noteworthy, the above asymptotic expression of $\tilde d(t)$ and $\tilde \lambda$ do not involve the details of the initial conditions.
If $r \geq r_\rmc$, the growth of $\tilde d(t)$ in Eq.~(\ref{eq:d_efg}) is sub-exponential and it can be described by a vanishing Lyapunov rate 
\begin{align}
\tilde \lambda = 0 \text{ if } r \geq r_\rmc\,,
\end{align}
which is the signature of non-chaotic dynamics.

\subsection{Spatially extended systems} \label{sec:spatial}
Here, we provide details on the continuous-time continuous-space formulation of the spatiotemporal spread of perturbation in extended systems subject to stochastic resetting. 
The state of the system is characterized by a spatially varying classical field which we denote by $\phi(t,x)$ to avoid confusion (until now $x$ denoted the state of the system, not the spatial position).
The dictionary with the discrete-time formulation on the lattice is similar to the one in~Sect.~\ref{sec:dico} where, now,
\begin{align}
    \begin{array}{rl}
    x_{n,\,i} & \longleftrightarrow \phi(t,x)\\
    \frac{x_{n,\,i-1} -2 \, x_{n,\,i} + x_{n,\,{i+1}}}{2} & \longleftrightarrow \nabla^2_x \phi(t,x)
    \end{array}
\end{align}

\subsubsection{Renewal formula for OTOCs}
One defines the classical OTOCs for the deterministic dynamics and the resetting dynamics as, respectively,
\begin{align}
    D(t,x) := \left| \frac{\delta \phi(t,x)}{\delta \phi(0,0)} \right|
\quad
\text{and}
\quad
    \tilde D(t,x) := \Big{\langle} \left| \frac{\delta \phi(t,x)}{\delta \phi(0,0)} \right| \Big{\rangle}_r\,.
\end{align}
Notably, these OTOCs obey a renewal equation that reads
\begin{align} \label{eq:renewal_cont_lattice}
\tilde D(t,x)  = r \int_0^t \rmd \tau \, \rme^{-r \tau} D(\tau,x) + \rme^{-r t} D(t,x)  \,.
\end{align}

\subsubsection{``Light-ray'' parametrization of the OTOC}
Here, we assume that the spatiotemporal dependence of the OTOC of the deterministic dynamics is captured by the following ansatz,
\begin{align} \label{eq:light_ray_param}
    D(t,x) = \exp\left( \lambda\left(\frac{|x|}{t} \right) t \right)\,,
\end{align}
where the even function $\lambda(v)$ is commonly referred to as a velocity-dependent Lyapunov exponent. It is assumed to be a continuous monotonically decreasing function with negative curvature, with  $\lambda(0) > 0$ and $\lambda(v)$ crosses zero at a finite velocity $v_\rmB$: $\lambda(v_\rmB) =0$.
In particular, on the light-ray traveling at the velocity $v_\rmB$, this yields
\begin{align}
  D(t, x= v_\rmB t) = 1\,.
\end{align}
Note that, although this is not explicit, the functional form in Eq.~(\ref{eq:light_ray_param}) supports a traveling wavefront at the butterfly velocity $v_\rmB$: take $x = v_\rmB t + z$ with $|z| \ll v_\rmB t$, expand $\lambda(v_\rmB + z/t)$ to first order in $z/t$, and obtain the following traveling wavefront
\begin{align}
    D(t,x = v_\rmB t + z)  \xrightarrow{t \to \infty}  \exp\left[-\frac{z}{v_\rmB} \lambda_\rmB \right]\,,
\end{align}
where the Lyapunov rate $\lambda_\rmB:=- v_\rmB\lambda'(v_\rmB) > 0$ sets the spatial extent of the exponentially-growing front.

\subsubsection{Light-cone propagation with stochastic resetting}
Here, we provide the derivations of the renormalized butterfly velocity that describes the effective speed of the ballistic spreading,
\begin{align}
    \tilde v_\rmB =
    \left\{
    \begin{array}{ll}
           \lambda^{-1}(r) & \text{ if } r < r_\rmc
     \\
    0 & \text{ if } r \geq r_\rmc
    \end{array}
\right., \text{ with } r_\rmc = \lambda(0) \,,\label{eq:vB_app}
\end{align}
where $\lambda^{-1}$ is the functional inverse of the velocity-dependent Lyapunov introduced in Eq.~(\ref{eq:light_ray_param}), as well as the renormalized Lyapunov that describes the spatial extent of the butterfly front,
\begin{align} \label{eq:lambdaB_app}
    \tilde \lambda_\rmB = - \tilde v_\rmB  \, \lambda'(\tilde v_\rmB)\,. 
\end{align}
Quite intuitively, we find both the butterfly velocity and the Lyapunov rate to be reduced by stochastic resetting: 
\begin{align}
    \partial_r \tilde v_\rmB \leq 0 \text{  and  } \partial_r \tilde \lambda_\rmB \leq 0 \,.
\end{align}

For simplicity of the algebra, we work in the framework of continuous-time dynamics, but the corresponding discrete-time expressions can be readily obtained without conceptual overhead.
In practice, let us check that there is indeed a wavefront traveling at the velocity $\tilde v_\rmB = \lambda^{-1}(r)$. Let us first treat the case $\tilde{v}_\rmB > 0 $, when $r < r_\rmc$.
The case of $\tilde {v}_{\rmB} = 0$, when $r \geq r_\rmc$, will be treated below.
We explore the vicinity of the light ray at $\tilde v_\rmB
$ by working with the parametrization $x = \tilde v_\rmB t + z$ and considering asymptotic times when $|z| \ll \tilde{v}_\rmB t $. Using the renewal equation (\ref{eq:renewal_cont_lattice}) and the shorthand notation $\tilde \lambda(v) := \lambda(v) - r$, we have
\begin{widetext}
\begin{align}
    \tilde D(t, x = \tilde v_\rmB t + z) &= r \int_0^t  \rmd \tau \exp\left[ \tilde \lambda\left(\frac{\tilde v_\rmB t + z}{\tau}\right) \tau \right]
    +  \exp\left[ \tilde  \lambda\left(\frac{\tilde v_\rmB t + z}{t}\right) t \right] \label{eq:renewal_inter}\\
&\simeq \exp\left[ z \tilde\lambda'\left( \tilde v_\rmB \right)\right] \left( 1+  r\exp\left[ -
z \tilde\lambda'\left( \tilde v_\rmB \right)\right] \int_0^t  \rmd \tau \exp\left[ \tilde \lambda\left(\tilde v_\rmB \frac{t}{\tau}\right) \tau + z \tilde\lambda'\left( \tilde v_\rmB \frac{t}{\tau} \right) \right]
      \right) \,,
\end{align}
\end{widetext}
where, in the second line, we expanded to first order in $z$ and used $\lambda(v_\rmB) = 0$.
Performing the change of variable $\tau \mapsto v:= \tilde v_\rmB {t}/{\tau}$ in the integral, we get
\begin{widetext}
\begin{align}
\tilde D(t, x = \tilde v_\rmB t + z) 
& \simeq\exp\left[ z \tilde\lambda'\left( \tilde v_\rmB \right)\right] 
 \left( 1+   r t \tilde v_\rmB \exp\left[ -z \tilde\lambda'\left( \tilde v_\rmB \right)\right] 
\int_{\tilde v_\rmB}^\infty   \frac{\rmd v}{v^2} \exp\left[  \tilde v_\rmB \frac{\tilde \lambda(v) }{v} t + z  \tilde\lambda'(v))\right]
      \right)\,.
\end{align}
\end{widetext}
Let us now take the limit $t\to\infty$. Note that the term $\tilde \lambda(v)$ in the integrand above is always negative, since $\tilde \lambda(v) := \lambda(v) - r$ crosses zero at $v = \tilde v_\rmB$.
Consequently, the term $\lambda(v)/v$ can only contribute to the integral close to $v \approx \tilde v_\rmB$ (\textit{i.e.} $\tau \approx t$). Let us therefore expand the integrand as
\begin{align}
    \exp\left[  \tilde v_\rmB \frac{\tilde \lambda(v) }{v} t + z  \tilde\lambda'(v) \right] 
  &\simeq
 \exp\left[   \tilde \lambda'\left(\tilde v_\rmB\right) \left\{ (v-\tilde v_\rmB) t + z \right\}\right] \,.
\end{align}
Performing the integral
\begin{align}
    \int_{\tilde v_\rmB}^\infty   \frac{\rmd v}{v^2}  \exp\left[   \tilde \lambda'\left(\tilde v_\rmB\right) (v-\tilde v_\rmB) t\right]  = - \frac{1}{t} \frac{1}{\lambda'(\tilde v_\rmB)}\,,
\end{align}
we obtain
\begin{align}
\lim\limits_{t\to\infty}  \tilde D(t, x = \tilde v_\rmB t + z) 
& =   \left( 1 -   r \frac{\tilde v_\rmB}{\lambda'(\tilde v_\rmB)}
      \right)
      \exp\left[ z \tilde\lambda'\left( \tilde v_\rmB \right)\right] \nonumber
      \\
   &  = \left( 1+   \frac{r}{ \tilde \lambda_\rmB} \right) \exp\left( - \frac{z}{\tilde v_\rmB} \tilde \lambda_\rmB \right)\,,
\end{align}
where we introduced  $\tilde \lambda_\rmB := - \tilde v_\rmB  \, \lambda'(\tilde v_\rmB) $.
This concludes the proof of the existence of a wavefront traveling ballistically with an effective butterfly velocity $\tilde v_\rmB$ and with an effective Lyapunov rate $\tilde \lambda_\rmB$ as given by Eq.~(\ref{eq:vB_app}) and (\ref{eq:lambdaB_app}), respectively.
In the particular case of a velocity-dependent Lyapunov of the form 
\begin{align} \label{eq:particular}
    \lambda(v) = \frac{\lambda_\rmB}{\alpha} \left[ 1-\left(\frac{v}{v_\rmB}\right)^\alpha \right] \mbox{ for } v>0
\end{align} 
with $\alpha \geq 1$, this yields 
\begin{align}
   \left\{
   \begin{array}{ll}
        \tilde v_\rmB &= v_\rmB \left(1 - \frac{r}{\lambda_\rmB} \right)^{1/\alpha} \\
    \tilde \lambda_\rmB &= \lambda_\rmB- r 
   \end{array}
   \right. \text{ for } r < r_\rmc = \lambda_\rmB\,,
\end{align}
and $\tilde v_\rmB = \tilde \lambda_\rmB = 0$ for $r > \lambda_\rmB$.

Let us now address the case $\tilde v_\rmB =0$, when $r \geq r_\rmc$. In the renewal Eq.~(\ref{eq:renewal_inter}), this case corresponds to $\tilde \lambda(v) < 0$ for all $v$. Hence, the renewal equation directly yields the stationary OTOC in terms of the Laplace transform
\begin{align}
   \tilde D_{\rm st}(z) := \lim\limits_{t\to\infty}  \tilde D(t,z) &=
   r \int_0^\infty \rmd\tau \exp \left[ \lambda(|z|/\tau) \tau -r \,\tau \right]\,,
\end{align}
which is maximum at $z=0$, where $\tilde D_{\rm st}(z=0) = r/(r- r_\rmc)$. This saturation value of the OTOC may be interpreted as a quantifier of the volume of phase space that is effectively contributing to the dynamics. Hence, it is a signature of the loss of ergodicity at $r \geq r_\rmc$. As an illustration, let us consider the particular velocity-dependent Lyapunov $\lambda(v)$ of Eq.~(\ref{eq:particular}) for $\alpha = 1$ and $\alpha = 2$.
This yields
\begin{align}
\tilde D_{\rm st}(z) = \frac{r}{r-r_\rmc} \exp\left(-|z|/\xi\right) \text{ 
 for } \alpha = 1\,,
\end{align}
with the constant localization length $\xi = v_\rmB/\lambda_\rmB$,
and
\begin{align}
\tilde D_{\rm st}(z) = \frac{r}{r-r_\rmc}
     \frac{|z|}{\xi}\ K_1\left( |z|/\xi \right)  &  \text{  for } \alpha = 2\,,
   \end{align}
where $K_1$ is the modified Bessel function of the second kind and the localization length $\xi = v_\rmB/\sqrt{2\lambda_\rmB (r-r_\rmc)}$ diverges at the transition.
Using the asymptotic form $K_1(z) \sim \sqrt{\pi/2z} \exp(-z)$ at large $z$, we find that both expressions above describe OTOCs that are exponentially localized around the perturbation site at $z=0$.

\bibliography{references}

\end{document}